# Ab Initio Modeling of Phonon-Assisted Relaxation of Electrons and Excitons in Semiconductor Nanocrystals for Multiexciton Generation

Taofang Zeng[§,¶,*], Yi He[§]

[§]Department of Mechanical Engineering, Massachusetts Institute of Technology, 77 Massachusetts Avenue, Cambridge, MA 02139

[¶]School of Energy and Power Engineering, Changsha University of Science and Technology, China

Electron–phonon and exciton–phonon interactions in nanoclusters are formulated and computed under the framework of GW-BSE (Bethe-Salpeter equation) approach. The phonon effect is modeled with the two-particle representation for the first time. The nonradiative relaxation rates of electrons and excitons are calculated. It is uncovered that both single-phonon relaxation and multiple-phonon relaxation are significant in nanocrystals, and correspond to two types of physical processes that have totally different spectral lineshapes. Furthermore, the multiple-phonon relaxation always occurs and its rates are comparable to the corresponding single-phonon relaxation rates for both electrons and excitons in the system studied (Si46). The inelastic scattering rates of electrons and excitons are also calculated based on many-body Green's function theory. For the electronic states in Si46, the inelastic scattering decay is predicted to be a primary decay mechanism for multiexciton relaxation, and nonradiative relaxation rates are larger than inelastic scattering rates for most excitonic states in Si46.
.

# 1. INTRODUCTION

In a typical photoexcitation process, absorption of sufficiently high energy photon in semiconductors initially promotes an electron from the valence band to the conduction band, and creates an electron-hole pair or quasiparticle or hot carrier. Hot carriers could be cooled quickly by losing energy through interacting with the host matrix or the lattice phonons. Because of strong electrons and carriers' interaction in nanocrystals (NCs), hot electrons in conduction band or holes in valence band also undergo inelastic scattering and promote another electron in the valence band to conduction band, in a way similar to impact ionization. This is the so-called multiplication (CM) or multiexciton generation (MEG). The thus produced multiexciton including biexciton, two pairs of electron-hole, or a pair of quasiparticle, and trion, can relax by Auger recombination among biexcitons and between trions and electron/hole back into single excitons.

CM can be used to enhance solar energy efficiency, and has been a topic under intense investigation for the development of the next-generation solar cell technology [1-6]. Theoretically, CM should be able to raise the Shockley-Queisser single-junction solar cell efficiency limit to around 44%. However, the actual improvement of the performance of solar cell thus far has been fairly limited. The great gap between what is obtained by experiments and what is expected by theory means that our understanding of the MEG in semiconductor NCs are far from complete.

The rate of CM or an MEG process is determined by the rate that an exciton transfers its energy partially to other electrons by generating another exciton in the same system. In other words, the rate of MEG is essentially the excitonic inelastic relaxation rate, or the inverse of the excitonic inelastic lifetime. For the purpose of comparison, in the following text the rate of the MEG process is also taken as the inelastic scattering rate of multiexciton.

It has long been known that to be an efficient solar cell candidate, the nanocrystals' MEG rate should be faster than the cooling rates of all other channels. More specifically, Spoor and coworkers[7] pointed out that the quantum yield for charge carrier photo-generation is the net result of the competitive relaxation of a hot electron−hole pair via CM and cooling by phonon emission.

In a series of experimental and mechanistic investigations, Spoor and coworkers[7-9] showed that both the CM rate and the phonon emission rate are energy-dependent, and the CM rate constant is about 0.3~0.7 $ps^{-1}$, and the phonon emission rate is about 147 ~174 $ps^{-1}$ for PbSe NCs and the photoexcitation energy range studied. Kambhampati and coworkers[10-11] proposed multiple possible pathways including phonon, ligand and Auger for exciton relaxation, and postulated that the difference between these relative contributions for relaxation yielded divergence between experimental results. They deduced a weak exciton-phonon coupling, and found that there was no phonon bottleneck for their experimental NCs, which led to a conclusion that phonon emission was of minor importance in controlling hot exciton dynamics. This is obviously in opposite to the results in references 7-9.

Another earlier experiment showed that the initial relaxation time of the 1P to 1S electronic states of CdSe nanocrystals was about 1ps, and following the initial relaxation, a slower cooling process has a relaxation time of 200ps[12]. They attributed the fast intraband relaxation to an electron-hole Auger process, and speculated that the slow relaxation was due to phonon bottleneck, a slow relaxation by phonons. Bonn and coworkers[13] measured the decay of exciton in InAs NCs, and found that the electron intraband relaxation from the 1Pe to 1Se level occurs on a time scale of 0.8 ps for particles of 2.2 nm radius. They also measured the Auger recombination (AR) rate of biexciton, and the AR rate was between 10ps to 40 ps.

Klimov and coworkers proposed an earlier two – competing energy relaxation mechanism for carrier multiplication to describe their experiements[14]. They suggested that the CM efficiency is set by two competing relaxation pathways of impact ionization-like scattering and non-CM relaxation due to phonon emission. Meanwhile they also attributed the cooling pathway of carriers as an Auger recombination (AR) process, and found out AR time constant between 10 to 1000ps based on various experimental data [3,15-16]. Another recent study reveals that triexciton emission in CdSe NCs is dominated by band-edge $1Se1S_{3/2}$ recombination and the triexciton lifetime is about 900 ps [17].

It can be seen that the above mentioned experimental results are distinct, and the researchers do not even agree to the exciton cooling pathways, not to mention the contribution of phonon emission to exciton relaxation. It must be pointed out that all these experiments have not directly measured the phonon signal. It was just recently that Cundiff and coworkers[18] reported the first direct probe of exciton-phonon coupling in colloidal quantum dots. They suggested the longitudinal-optical phonon coupling as a major factor in spectral diffusion on femtosecond time-scale.

As more new experiments provide direct evidence that phonon emission is a main cooling pathway, yet some researchers suggested that elastic Auger process plays an important role in carrier cooling in some publications.

The plausibility of the cooling channels and their relative contributions to exciton relaxation in semiconductor nanocrystals has also been reflected in theoretical studies. By assuming the same intraband relaxation lifetime of 0.5 ps for phonons, Delerue and coworkers [19] fitted five different sets of CM experimental data for PbSe NCs by using tight binding theory and impact ionization theory. Adopting the same computation and experimental methods for HgTe NCs, they found a phonon lifetime of 6 ps, and a biexciton lifetime of 49 ps [20]. Liu and coworkers calculated MEG rate in CdSe NCs by using an atomistic pseudopotential approach and first-order perturbation theory, an incoherent relaxation assumption, and obtained a relatively low MEG rate[21]. By solving the Markovian quantum master equation, Azizi and Machnikowski studied the impact of dissipation on the evolution of the single exciton and biexciton occupations, and found that in a certain range of parameters, the MEG rate is fast (on picosecond timescale) and the following decay is much slower [22]. It can be seen that the two results for MEG rates from ref. 21 and 22 are in direct contradiction. This clearly shows the importance of selection of modeling approach.

It is evident that the conclusions and arguments about the contribution of phonon emission for carrier relaxation in nano-clusters are remarkably divergent. In fact, Efros et al. [6] recently pointed out that it was lack of a reliable description of the major mechanisms of the carrier thermalization in NCs. In their recent articles [23-24], they reported experimental and computational results for dynamics of intraband and interband Auger processes in NCs. Their computation is based on an eight-band **k•p** method with standard time-dependent perturbation theory. They argued that "normal" cooling through the emission of phonon is impossible because there are no phonon modes of sufficiently high energy to bridge the separation between the conduction band levels. Instead, hot electron cooling was mainly due to its Auger coupling with the hole, where the hot electron transfers the relaxation energy to a valence band hole, and the hole then rapidly cools down to the top of the valence band by emission of phonons [23-25]. This is consistent with references 12-13. Their speculation on the role of Auger and phonon cooling, however, is in contrary to the newest experimental results[7,18], and is also inconsistent with the computational results with the first-principle based modeling such as the time-dependent density function theory (TDDFT) computation. The latter two have uncovered a major effect by direct electron-phonon interaction.

The phenomenological and semi-classical modeling for contributions of phonon emissions to MEG is inconclusive, and a rigorous and accurate approach that is not empirically based is desirable. An ideal approach would be first-principle based and capable of dealing with excited electrons and phonon emissions for MEG dynamics. One of such approaches is the TDDFT computation. Prezhdo and coworkers have applied this approach to model a suite of NCs and studied the MEG dynamics including light-matter interaction, phonon-induced dephasing and electron relaxation, and reverse Auger recombination[5,26-28]. TDDFT, however, can only be used to compute very small nanoclusters and for a short duration of the dynamic processes because of the massive CPU demanding by TDDFT. Actually, it can be seen that their modeling has not reached steady state. For example, the SE (single exciton) population still rapidly decays after duration of 3 ps of modeling as shown in ref. 5.

Therefore it should be cautioned when the results are used to evaluate the performance of an MEG – based device, the net effect of photoexcitation under steady state. Additionally, in the TDDFT computation single-electron KS orbitals are generally adopted. TDDFT is actually TDKS under the framework of one-particle representation, and TDKS often fails for charge-transfer excited states, multiple excitations and avoided crossings [29-31].

An alternative approach to model the effect of phonon emission to MEG dynamics is the *ab initio* GW-Bethe-Salpeter equation (BSE). The GW can be used to accurately calculate electronic structures, and the BSE explicitly includes the exchange and dynamic screened Coulomb interaction between two particles, and is the standard for the simulation of excitons in bulk materials. Louie and coworkers have applied the approach to model the effects of doping on excitons and carrier lifetimes [32], hot carrier thermalization in bulk silicon under sunlight illumination [33], and singlet fission in solid pentacene [34]. Their modeling agreed well with experiments available. Based on GW-BSE, Rabani and coworkers modelled MEG in semiconductor NCs with input parameters from experimental phonon self-energy [35], Auger recombination in semiconductor NCs [36], and MEG efficiency in nanorods within the static screening approximation and with semi-empirical dielectric constant input [37].

In our previous research, we developed a GW-BSE approach for directly modeling of energies and lifetimes of electrons and excitons in small silicon nanoclusters and quasiparticle lifetimes in magnesium clusters [38-39]. In the modeling, an initially excited electron naturally transfers part of its energy above the band gap to another electron and promotes it to the conduction band, which produces another exciton, and thus forming a biexciton. Approximations including the Tamm-Dancoff are made to enable calculability of BSE without use of any input parameter. Self-consistency of the G function and reduced polarizability are considered. Our modeling for silicon nanoclusters has been verified by comparing the computed ionization potential and optical absorption spectra with experiments.

Reviewing the existing literatures, one can find that the controversy about contribution of phonon emission to MEG in NCs still remains, though more new experiments confirm a fast phonon-dominant pathway. A theory or simulation that can clarify the phonon related processes in NCs is highly needed.

Nevertheless electron-phonon and exciton-phonon have been extensively studied in an ab initio manner. Some related literatures are reviewed herein. Piryatinski and Velizhanin developed an exciton scattering model for the CM dynamics [40]. Yet the model has found any application in modeling. By assuming that the eigenstates of excitons are a linear combination of electron-hole pairs, Marini derived equations for temperature dependence of the excitonic energies, and nonradiative excitonic lifetime, which were used to explain optical absorption spectra of bulk silicon and hexagonal boron nitride [41]. Marini and his colleagues later on developed an electron-phonon interaction theory based on many-body perturbation formalism [42]. The complication of the formula, however, makes it too difficult to execute numerical computation or modeling. Recently, Antonius and Louie developed theory for the exciton-phonon coupling [43], whose application in modeling has not been reported yet.

Some review articles have well documented the progresses in the field of electron-phonon interaction and phonon-assisted dynamics of electrons and excitons in NCs[44-45]. Summarizing the above literatures, one can find that ab initio (without any semiempirical parameter input), GW-BSE based formulas and corresponding approach that are computationally executable are yet to be developed for accurate and feasible calculations for electron-phonon interaction and exciton-phonon interaction in semiconductor NCs.

The goal of this research is to contribute such an effort to lay out a fundamental understanding for phonon effect on exciton relaxation, more specifically for singlet exciton relaxation with the presence of phonon emission. Based on our previous GW-BSE modeling for quasiparticle and exciton lifetimes in nanostructures, electron – phonon and exciton – phonon interactions are formulated and modeled under the framework of GW-BSE approach in the investigation. It is the first time that phonon effect is computed with the two-particle representation, i.e. the excitonic states are treated as a combination of electrons and holes. It finds that both single-phonon and multiple phonon effects are major pathways for exciton and MEG dynamics in the semiconductor NCs studied. It also shows that the inelastic scattering decay is a major decay mechanism, and the nonradiative relaxation rates are larger than the inelastic scattering rates for most excitonic states in Si46.

## 2. METHODOLOGY

### 2.1. Many-Body Green's Function Theory and Inelastic Scattering Rates

Formulas and approaches for the computation of quasiparticle lifetimes and inelastic scattering (carrier multiplication) rates in nano-clusters have been reported in our previous articles [38,39]. For the completeness of the present article, a brief description for many-body Green's function theory is presented herein.

#### 2.1.1. Quasiparticle excitation:

The electronic energies of a many-body system can be obtained by solving the quasiparticle (QP) equation

$$(T + V_{ext} + V_H)\varphi_i(\mathbf{r}) + \int d\mathbf{r}' \Sigma_{xc}(\mathbf{r}, \mathbf{r}'; E_i)\varphi_i(\mathbf{r}') = E_i \varphi_i(\mathbf{r}) \qquad (1)$$

where $T$ is the kinetic energy operator, $V_{ext}$ the external potential, $V_H$ the Hartree potential, $E_i$ and $\varphi_i$ the energy and wavefunction of the $i$th QP, and $\Sigma_{xc}(\mathbf{r},\mathbf{r}';E_i)$ the exchange-correlation self-energy operator. The QP equation is solved based on the results of the density functional theory (DFT)

$$(T+V_{ext}+V_H)\varphi_i(\mathbf{r})+V_{xc}(\mathbf{r})\varphi_i(\mathbf{r})=\varepsilon_i\varphi_i(\mathbf{r}) \qquad (2)$$

where $\varepsilon_i$ and $\varphi_i$ are the eigenvalue and eigenfunction of the $i$th Kohn-Sham (KS) particle respectively, and $V_{xc}(\mathbf{r})$ the exchange-correlation potential. With the assumption that the KS eigenfunctions agree well with the QP wavefunctions in most cases, QP energies are usually solved with perturbative method to the first order

$$\langle\varphi_i|\Sigma_{xc}(\mathbf{r},\mathbf{r}';E_i)|\varphi_i\rangle-\langle\varphi_i|V_{xc}(\mathbf{r})|\varphi_i\rangle=E_i-\varepsilon_i, \qquad (3)$$

According to Hedin's equations,[46] $\Sigma_{xc}=i\hbar GW\Gamma$, where $\Gamma$ is the vertex function and $G$ is the one-particle Green's function

$$G(\mathbf{r},\mathbf{r}';E)=\sum_n\frac{\varphi_n(\mathbf{r})\varphi_n(\mathbf{r}')}{E-E_n+i\eta_n 0^+}. \qquad (4)$$

The coefficient $\eta_n$ is $+1$ for unoccupied states and $-1$ for occupied states. $W$ is the screened Coulomb interaction which can be written as

$$W=V+V\Pi V, \qquad (5)$$

where $V(\mathbf{r},\mathbf{r}')$ is the Coulomb interaction, and $\Pi(\mathbf{r},\mathbf{r}';E)$ is the reducible polarizability and can be expressed as the summation of well-defined resonant modes.[47]

$$\Pi(\mathbf{r},\mathbf{r}';E)=2\sum_s\rho_s(\mathbf{r})\rho_s^*(\mathbf{r}')\left[\frac{1}{E-(\omega_s-i0^+)}-\frac{1}{E+(\omega_s-i0^+)}\right], \qquad (6)$$

where

$$\rho_s(\mathbf{r})=\sum_{v,c}R_s^{v,c}\varphi_v^*(\mathbf{r})\varphi_c(\mathbf{r}) \qquad (7)$$

is the particle-hole amplitude for the $s$th particle-hole excitations. The eigenvectors $\mathbf{R}_s^{v,c}$ and eigenvalues $\omega_s$ are obtained by solving the equation $\Pi=P+PV\Pi$, where $P=-iGG\Gamma$ is the irreducible polarizability. Both $\Sigma_{xc}$ and $\Pi$ (or W) include the vertex function $\Gamma$. It has been shown that a consistent choice of $\Gamma$ is necessary for the QP calculation. In this paper, $\Gamma$ is obtained by solving the equation $\Gamma = 1 + \left(\partial \Sigma^0 / \partial G\right) G G \Gamma$ in the framework of the local density approximation (LDA), which is equivalent to the time-dependent LDA (TDLDA) for $\Pi$ and $GW\Gamma$ for $\Sigma_{xc}$. The self-energy term $\langle \varphi_i | \Sigma_{xc} | \varphi_i \rangle$ in Eq. (3) can be separated into an energy-independent exchange part $\langle \varphi_i | \Sigma_x | \varphi_i \rangle$ and an energy-dependent correlation part $\langle \varphi_i | \Sigma_c | \varphi_i \rangle$. The latter is evaluated as[48]

$$\langle \varphi_i | \Sigma_c(\mathbf{r},\mathbf{r}';E) | \varphi_i \rangle = \sum_n \sum_s \frac{a_{n,s,i}}{E - E_n - \omega_s \eta_n}, \text{ where } a_{n,s,i} \text{ equals}$$

$2\langle \varphi_i \varphi_n | (V + f_{xc}) | \rho_s \rangle \langle \rho_s | V | \varphi_i \varphi_n \rangle$ in the $GW\Gamma$ implementation.

The imaginary parts of the QP energies can be obtained by applying analytical continuation of $\Sigma_{xc}(\mathbf{r},\mathbf{r}';E)$ in the complex energy plane, and the complex QP energy $E_i - i\eta_i \gamma_i$ is calculated by solving a complex equation set numerically

$$\text{Re}\langle \varphi_i | \Sigma_{xc}(E_i - i\eta_i \gamma_i) | \varphi_i \rangle - \langle \varphi_i | V_{xc} | \varphi_i \rangle = E_i - \varepsilon_i \quad (8a)$$

$$\left|\text{Im}\langle \varphi_i | \Sigma_{xc}(E_i - i\eta_i \gamma_i) | \varphi_i \rangle\right| = \gamma_i \quad (8b)$$

where $\gamma_i$ is the inelastic scattering rate of the *i*th QP because of electron-electron interaction.

### 2.1.2. Electron-hole interaction and the Bethe-Salpeter equation:

An excitonic state of a system with *N* electrons essentially involves two particles, which can be investigated by the Bethe-Salpeter equation (BSE)[49-51]

$$L(1,2;1',2') = G(1,2')G(2,1') + \int d(33'44') G(1,3)G(3',1') \Xi(3,4';3',4) L(4,2;4',2'), \quad (9)$$

where $L(1,2;1',2')$ is the two-particle correlation function. In Eq. (9) a integer label is assigned to a set of space, spin and time variables, namely $(1) = (\mathbf{x}_1, t_1) = (\mathbf{r}_1, \sigma_1, t_1)$. The integral kernel $\Xi$ can be approximated as[51]

$$\Xi(3,4';3',4) \approx -i\delta(3,3')\delta(4^+,3')V(3,4) + i\delta(3,4)\delta(3',4')W(3^+,3') \quad (10)$$

Thus the BSE Eq. (9) can be converted to a complex eigenvalue problem

$$\left[(E_c - i\gamma_c) - (E_v + i\gamma_v)\right] A^q_{vc} + \sum_{v',c'} A^q_{v'c'} \left(K^x_{vcv'c'} + K^d_{vcv'c'}\right) = (\Omega_q - i\Gamma_q) A^q_{vc} \qquad (11)$$

Where $\gamma_c$ and $\gamma_v$ are damping rates of electrons and holes respectively, and $\Omega_q$ and $\Gamma_q$ on the right hand side of Eq. (11) are the excitation energy and the relaxation rate of the $q$th exciton respectively. In this paper, only singlet excitations are considered, and thus the exchange term is $K^x_{vcv'c'} = 2\langle \varphi_v \varphi_c | V | \varphi_{v'} \varphi_{c'} \rangle$. The direct interaction term $K^d_{vcv'c'}$ can be calculated as

$$\begin{aligned} K^d_{vcv'c'} = &-\langle \varphi_v \varphi_{v'} | V | \varphi_c \varphi_{c'} \rangle \\ &-\sum_s \left\{ \begin{array}{l} \left( \dfrac{1}{\Omega_q - i\Gamma_q - \omega_s - (E_{c'} - E_v)} + \dfrac{1}{\Omega_q - i\Gamma_q - \omega_s - (E_c - E_{v'})} \right) \\ \times \left( \langle \varphi_v \varphi_{v'} | V | \rho_s \rangle \langle \rho_s | (V + f_{xc}) | \varphi_c \varphi_{c'} \rangle + \langle \varphi_v \varphi_{v'} | (V + f_{xc}) | \rho_s \rangle \langle \rho_s | V | \varphi_c \varphi_{c'} \rangle \right) \end{array} \right\} \end{aligned} \qquad (12)$$

### 2.1.3. Approximation and solutions:

Actually Eq. (11) explicitly includes four terms related to the decay of the exciton, which are illustrated by the Feynman diagrams in Fig.1.

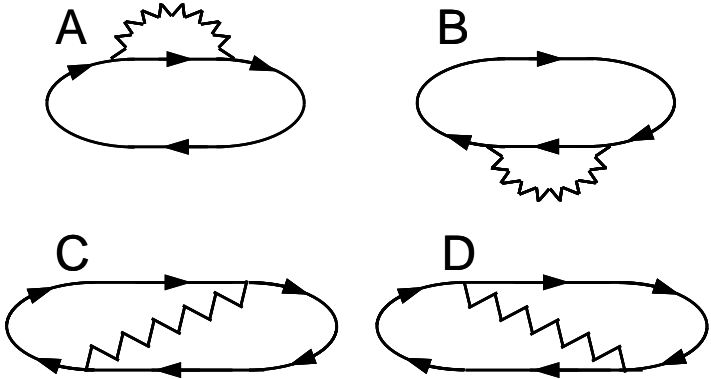

Fig. 1. Feynman diagrams of terms in Eq. (11) related to the decay of particle-hole excitations. Arrowed lines are Green's functions. Wiggled lines are screened interactions. Diagrams A and B correspond to the diagonal elements in Eq. (11). Diagrams C and D denote the screened particle-hole interaction in Eq. (11).

In Eq. (11) only the resonant part in included, while the antiresonant part is neglected. This is the Tamm-Dancoff approximation, whose effect on excitonic energies was found to be negligible. However, it is unfeasible to directly solve Eq. (11), which we call dynamic BSE, since the matrix on the left hand side is explicitly dependent on the eigenvalues $\Omega_q$ to be solved.

We have demonstrated that an approximate method that only takes into account the first two diagrams in Fig. 1 can be used to solve Eq. (11), which was simplified to a computable frequency-independent eigen-problem[39]. In the computation, it was found that partial self-consistence of G and Π provides stable numerical computation, where only complex eigenvalues are updated iteratively by the frequency-independent BSE matrix, with eigenvectors fixed to the TDLDA amplitudes. To further accelerate and stabilize the self-consistency procedure, an important and useful initial guess for the imaginary part of the excitation energy for a given exciton was made:

$$\Gamma_q = \sum_{vc} |R_{vc}^q|^2 (\gamma_c + \gamma_v), \qquad (13)$$

Where, $R_{vc}^q$ is the coefficient that defines the qth exciton in terms of an expansion of quasi-electron and quasi-hole. The above initial guess for the qth exciton's energy can be rationalized by the fact that the true excitonic eigenstates shall be close to the approximated eigenstates of excitons with a linear combination of electron-hole pairs as in ref. 41. The absorption spectra obtained by our approximation approach are found to be in good agreement with those obtained by the dynamic BSE (Eq. 11)[39], which are computed by using the results from the approximate method as the initial guess input. This validates our approximation for solving BSE.

**2.2. Electron-Phonon Interaction**

It is indispensible to study MEG dynamics through carriers' interaction with their host matrix by activating lattice vibration (phonon). Though there are extensive literatures on electron-phonon interaction[40-45], the formulas derived in those articles have yet found applications in numerical modeling for semiconductor NCs, probably due to their complications and coding infeasibility. It is thus essential to formulate the interactions that are feasible for programming and are coherent with the *ab initio* many-body Green's function theory. The electron – phonon and exciton – phonon interactions are formulated under the framework of GW-BSE approach as follows. It should be emphasized that the term "nonradiative relaxation" may stand for different processes under different circumstances. In this article, it is further categorized as a single-phonon relaxation process and a multi-phonon relaxation process of carriers.

The single-phonon process means that an electron makes a phonon-assisted transition from an electronic state transits to another electronic state, with the electron emitting or absorbing a phonon at the same time. In the field of physics, the theoretical work on this process originated from the investigation of the temperature-dependence of the optical gaps of bulk silicon and germanium. The process leads to the broadening of the electronic states in semiconductors. The formula associated with this process was first developed by Fan[52] in 1951, and the derived electron-phonon self-energy term is called the "Fan" term, where the imaginary part of the Fan term is exactly the single-phonon relaxation rate. Note that the temperature effect also manifests itself through the thermal expansion of the crystal lattice, and through a Debye–Waller term corresponding to the elastic interaction between electrons and phonons.[53] However, the thermal expansion and Debye-Waller terms only influence the real parts of the electronic energies, and make no contribution to their imaginary parts. This means that the two effects are irrelevant for describing the finite lifetimes of electronic states, and will not be covered in this study.

The multiple-phonon process means that an electron in one electronic state makes a transition to another electronic state, with the quantum numbers of several coupled phonon modes changed at the same time. In the field of chemistry, the theoretical work describing this electronic process originates from the investigation of radiationless transitions of electronic states in large molecules.[54] Usually the energy gap between the first two electronic states is so large that it cannot be matched just by the energy of one phonon. Therefore a process involving multiple phonons is the only possible relaxation mechanism. The formulas for the transition rates have been developed based on the perturbation theory. It should be pointed out that the anharmonic effect has to be taken into account for multiple-phonon processes, and such processes are attributed to the displacement of the potential energy surface during the electronic transitions.[55]

Physical and chemical researches have tackled the electron-phonon interaction from different aspects. The question is which aspect we shall follow for the electron-phonon interaction in a semiconductor NC. If a NC is more like a bulk material then we should focus on a single-phonon process, or if it is more like a molecule then we should focus on a multiple-phonon process. The question has not been addressed in existing literatures as far as can be found. In this study we show that both mechanisms should be included. Because a NC or a nano-cluster takes essentially the transition peculiarity in-between a bulk material and a molecule, it has all features of both matters, i.e. molecules and bulk materials.

The argument can be rationalized with the assistance of the schematic energy diagram for a semiconductor nano-cluster schematically shown in Fig. 2. In the cluster, an electron in the electronic state on the top can jump to those states right below it through a single-phonon process, as the energy gaps between these states and the top one are smaller than the energy of one phonon $\omega_k$. On the other hand, the electron in the electronic state on the top can also jump to those states far below it through a multiple-phonon process, since now the energy gaps between these states and the top one are so large that single-phonon process is prohibited. According to Fig. 2, we can find that the final states available for a single-phonon process are fewer than those for a multiple-phonon process. However, the single-phonon process is usually faster than the multiple-phonon process. Therefore the contributions of the single-phonon process and multiple-phonon process to the overall nonradiative relaxation rate of a high energy state could be comparable in magnitude. This means that both the single-phonon and the multiple-phonon mechanisms are essential for the calculation of nonradiative relaxation rates of electronic states in nanoclusters.

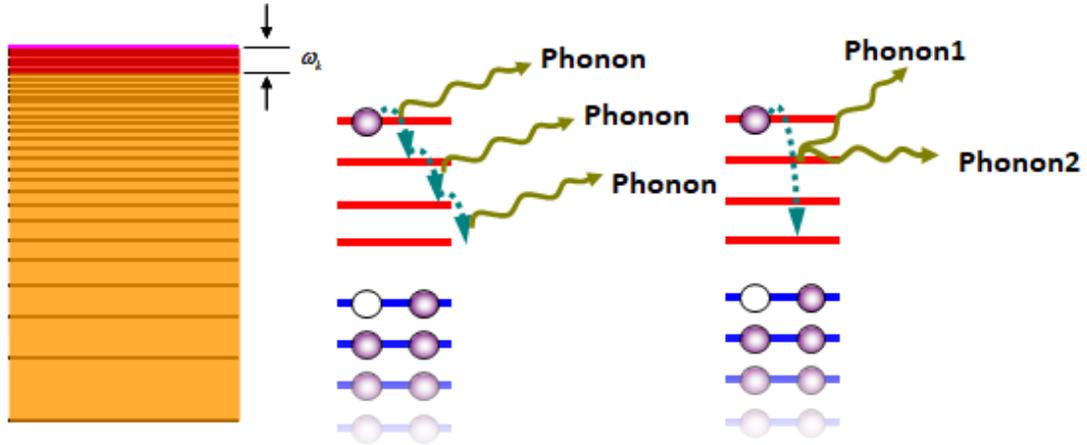

Fig.2 Schematic illustration of nonradiative relaxation processes and energy diagram of a semiconductor nano-cluster. In the left pane, the top electronic state (magenta one) can jump to the states right below it (red upper zone) through a single-phonon process (middle pane), or jump to the states far below it (orange lower zone) through a multiple-phonon process (right pane). The energy criterion to distinguish between the two mechanism is based on the single phonon frequency.

**2.2.1. One-particle representation vs. two-particle representation**

As it can be found, all existing simulations (computations) of electron-phonon interactions are conducted in the framework of a one-particle representation, and the nonradiative relaxation of an excited state is always treated as the decay of an electron or a hole. This is acceptable for some cases when electrons and holes themselves are essentially at one-particle states. However, treating excitons in a one-particle representation implies an independent-particle assumption, where excitonic states are written as $|i,j\rangle$, instead of a combination of $|i,j\rangle$. With this assumption, the decay of an excitonic state $|i,j\rangle$ does simplify to the decay of its electron component ($|i,j\rangle \rightarrow |i,j'\rangle$) or its hole component ($|i,j\rangle \rightarrow |i',j\rangle$).

The independent-particle assumption may hold for the first several excitonic states, classified according to their energies, because in most cases they can be approximated by $|i,j\rangle$. However, high-energy excitons shall always be expressed as the linear combination of $|i,j\rangle$, because the one-particle picture is expected to break down for these excitonic states. We thus propose treating all exciton-phonon interaction in the framework of two-particle presentation. Changing from the one-particle presentation to the two particle presentation will significantly affect the phonon-assisted relaxation rates of excitons, which is proportional to the scaling of the density of final states and the square of Coulomb coupling.[19, 56, 57] Usually the density of final states determines the available number of transition channel, the higher the final DOS, the higher the transition rates could be. As will be shown, the two-particle density of state (DOS) can be regarded as a convolution of the electron DOS and the hole DOS. The excitonic DOS is much higher, which contributes to the significant difference between the relaxation rates of excitons and those of electrons.

### 2.2.2. Adiabatic approximation

The Hamiltonian of a system composed of electrons and nuclei can be expressed as

$$H = T(\mathbf{r}) + T(\mathbf{Q}) + U(\mathbf{r},\mathbf{Q}), \tag{14}$$

where $\mathbf{r}$ and $\mathbf{Q}$ are the coordinates of electrons and nuclei, respectively, $T(\mathbf{r})$ and $T(\mathbf{Q})$ the kinetic energy operators of the electrons and nuclei, and $U(\mathbf{r},\mathbf{Q})$ is the total potential energy among all electrons and nuclei.

Within the framework of the Born-Oppenheimer approximation[58], the wavefunctions of electrons and nuclei are assumed to be independent. Thus the wavefunctions $\Phi_i(\mathbf{r},\mathbf{Q})$ and energies $E_i(\mathbf{Q})$ of electrons can be obtained for each nuclear configuration $\mathbf{Q}$ by solving the electronic Schrödinger equation

$$[T(\mathbf{r}) + U(\mathbf{r},\mathbf{Q})]\Phi_i(\mathbf{r},\mathbf{Q}) = E_i(\mathbf{Q})\Phi_i(\mathbf{r},\mathbf{Q}). \tag{15}$$

The wavefunction $\psi(\mathbf{r},\mathbf{Q})$ of the whole system can be expanded with $\Phi_i(\mathbf{r},\mathbf{Q})$ as the basis as

$$\psi(\mathbf{r},\mathbf{Q}) = \sum_i \chi_i(\mathbf{Q})\Phi_i(\mathbf{r},\mathbf{Q}). \tag{16}$$

Substituting this wavefunction into the Schrödinger equation of the system

$$H\psi(\mathbf{r},\mathbf{Q}) = [T(\mathbf{r}) + T(\mathbf{Q}) + U(\mathbf{r},\mathbf{Q})]\psi(\mathbf{r},\mathbf{Q}) = V\psi(\mathbf{r},\mathbf{Q}), \tag{17}$$

and then projecting both sides onto the electronic wavefunction $\Phi_j(\mathbf{q},\mathbf{Q})$, we will obtain a set of coupled equations for $\chi_i(\mathbf{Q})$

$$\sum_i H_{ij}(\mathbf{Q})\chi_i(\mathbf{Q}) = V\chi_j(\mathbf{Q}) \tag{18}$$

Where $\chi_i(Q)$ is nuclear wavefunction, $V$ is the energy of the state $\psi(\mathbf{q},\mathbf{Q})$, and $H_{ij}(\mathbf{Q})$ is the Hamiltonian

$$H_{ij}(\mathbf{Q}) = H_{ij}^0(\mathbf{Q}) + H_{ij}^1(\mathbf{Q}) \tag{19}$$

where the unperturbed and perturbation Hamiltonian are

$$H_{ij}^0(\mathbf{Q}) = \delta_{ij}\left[E_i(\mathbf{Q}) - \sum_k \frac{\hbar^2}{2M_k}\frac{\partial^2}{\partial Q_k^2}\right] \tag{20a}$$

$$H_{ij}^1(\mathbf{Q}) = -\sum_k \frac{\hbar^2}{2M_k}\left(2\left\langle\Phi_j\left|\frac{\partial}{\partial Q_k}\right|\Phi_i\right\rangle\frac{\partial}{\partial Q_k} + \left\langle\Phi_j\left|\frac{\partial^2}{\partial Q_k^2}\right|\Phi_i\right\rangle\right) \tag{20b}$$

where $M_k$ are the masses of the normal coordinates $Q_k$.

Within the framework of the adiabatic approximation, the non-adiabatic coupling term $H_{ij}^1(\mathbf{Q})$ is neglected.[54] Therefore the Hamiltonian matrix becomes diagonal, and Eq. (18) is simplified to

$$\left(E_i(\mathbf{Q}) - \sum_k \frac{\hbar^2}{2M_k}\frac{\partial^2}{\partial Q_k^2}\right)\chi_i(\mathbf{Q}) = V\chi_i(\mathbf{Q}) \tag{21}$$

which implies the nuclei move on the adiabatic potential energy surface (PES) $E_i(\mathbf{Q})$. The nuclear wavefunctions $\chi_i(\mathbf{Q})$ can be obtained by solving Eq. (21).

Within the harmonic approximation, all anharmonic effects are neglected. Thus $E_i(\mathbf{Q})$ can be expressed as the linear combination of linear and quadratic terms,

$$E_i(\mathbf{Q}) = \sum_k \alpha_k^i Q_k + \sum_k \beta_k^i Q_k^2 + \sum_{k,l} \gamma_{k,l}^i Q_k Q_l \tag{22}$$

where $\alpha_k^i$, $\beta_k^i$, $\gamma_{k,l}^i$ are coefficients. By choosing the equilibrium position as the origin $\mathbf{Q}_0$, $\alpha_k^i$ can be eliminated. For normal coordinates $Q_k$, bilinear terms vanish and $\gamma_{k,l}^i = 0$. Therefore Eq. (21) yields $k$ independent one-dimensional harmonic-oscillator equations, which have analytical solutions $\theta_{i,v_k}(Q_k)$ in which $v_k$ are the quantum numbers. Then the nuclear wavefunction $\chi_{i,v}(\mathbf{Q})$ is expressed as

$$\chi_{i,v}(\mathbf{Q}) = \prod_k \theta_{i,v_k}(Q_k), \tag{23}$$

where $v = (v_1, v_2, \ldots, v_k)$.

### 2.2.3. Perturbation theory

Within the framework of the perturbation approximation, the nonradiative transition rate between any two adiabatic states $\chi_{i,v'}(\mathbf{Q})\Phi_i(\mathbf{r},\mathbf{Q})$ and $\chi_{j,v''}(\mathbf{Q})\Phi_j(\mathbf{r},\mathbf{Q})$ with energies $V_{i,v'}$ and $V_{j,v''}$ can be calculated with the Fermi golden rule by taking $H_{ij}^1(\mathbf{Q})$ as the perturbation Hamiltonian,[59,60]

$$W_{i \to j} = \frac{2\pi}{\hbar} \sum_{v',v''} P_{v'} \left| \left\langle \chi_{j,v''} \left| H_{ij}^1 \right| \chi_{i,v'} \right\rangle \right|^2 \delta(V_{j,v''} - V_{i,v'}) \tag{24}$$

where the summation is over all initial vibronic states $v'$ weighted by the Boltzmann factor $P_{v'}$, and all final vibrational states $v''$. The perturbation term is

$$\left\langle \chi_{j,v''} \left| H_{ij}^1 \right| \chi_{i,v'} \right\rangle = -\sum_k \frac{\hbar^2}{M_k} \left\langle \Phi_j \chi_{j,v''} \left| \frac{\partial \Phi_i}{\partial Q_k} \frac{\partial \chi_{i,v'}}{\partial Q_k} \right\rangle - \sum_k \frac{\hbar^2}{2M_k} \left\langle \Phi_j \chi_{j,v''} \left| \chi_{i,v'} \frac{\partial \Phi_i^2}{\partial^2 Q_k} \right\rangle \right. \tag{25}$$

The second term in Eq (25) are usually neglected with the assumption that the electronic wavefunctions are slowly varying functions of the normal coordinates $Q_k$, and Eq (25) becomes[61,62]

$$\left\langle \chi_{j,v''} \left| H_{ij}^1 \right| \chi_{i,v'} \right\rangle = -\sum_k \frac{\hbar^2}{M_k} \left\langle \Phi_j \left| \frac{\partial \Phi_i}{\partial Q_k} \right\rangle \left\langle \chi_{j,v''} \left| \frac{\partial \chi_{i,v'}}{\partial Q_k} \right\rangle \right. \tag{26}$$

### 2.2.4. Single-phonon relaxation rates

The lineshape function is the crucial issue to evaluate relaxation rate $\Gamma_{i \to j}$ numerically, which is essentially determined by the underlying decay mechanisms. As shown in Fig.2, a high-energy exciton can decay through both single-phonon and multiple-phonon processes. Both processes should be considered. Here we propose an energy criterion to distinguish between the two processes, $\Delta E_{ij} < \hbar \omega_k$ for single-phonon relaxation, and $\Delta E_{ij} > \hbar \omega_k$ for multiple-phonon relaxation.

For a single-phonon process, the decay only occurs between two electronic states with an energy difference smaller than that of the $k^{\text{th}}$ phonon, $\hbar \omega_k$. In this case a Lorentzian lineshape is applied and Eq. (24) is simplified as

$$\gamma_{i \to j}^{SP} = \sum_k \frac{\hbar^2}{M_k} \left| C_k^{ij} \right|^2 \hbar \omega_k \left[ \frac{(n_k + 1) \gamma_i^{SP}}{\left( \Delta E_{ij} - \hbar \omega_k \right)^2 + \left( \gamma_i^{SP} \right)^2} + \frac{n_k \gamma_i^{SP}}{\left( \Delta E_{ij} + \hbar \omega_k \right)^2 + \left( \gamma_i^{SP} \right)^2} \right] \quad (27)$$

where $C_k^{i,j} = \left\langle \varphi_j \left| \partial / \partial Q_k \right| \varphi_i \right\rangle$, $\Delta E_{ij} = E_i - E_j$, and $\gamma_i^{SP}$ is the width of the Lorentzian function corresponding to the $i^{\text{th}}$ electron, which is exactly the single-phonon decay rate to be determined. Here $n_k$ is the average quantum number of the $k^{\text{th}}$ vibrational mode at the thermal equilibrium. $\gamma_{i \to j}^{SP}$ can be found as the imaginary part of the following self-energy term

$$\Sigma_{i \to j}^{sp} = \sum_k \frac{\hbar^2}{M_k} \left| C_k^{ij} \right|^2 \hbar \omega_k \left[ \frac{n_k + 1}{\Delta E_{ij} - \hbar \omega_k - i\gamma_i^{SP}} + \frac{n_k}{\Delta E_{ij} + \hbar \omega_k - i\gamma_i^{SP}} \right] \quad (28)$$

This term is close to the self-energy term in Ref. **52**, with a difference in coefficients arising from different perturbation mechanisms. Note that $\Sigma_{i \to j}^{sp}$ corresponds to a Feynman diagram similar to $\Sigma_{xc}$ in the *GW* case. Thus $\gamma_i^{SP} = \sum_{j, |E_j| < |E_i|} \gamma_{i \to j}^{SP}$ can be evaluated numerically in the same manner as that for the QP inelastic scattering rates in the *GW*Γ implementation[39].

### 2.2.5. Multiple-phonon relaxation rates

The multiple-phonon decay process is more complicated and can only be treated properly by including anharmonic effects. Since a small cluster can be regarded as a poly-atomic molecule, the dominant anharmonic effect is attributed to the displacement of the potential energy surface for different electronic states.[63] By adopting the displaced potential surface approximation, the normal coordinates $Q_k$ and their masses $M_k$ and frequencies $\omega_k$ are assumed to be constant for all electronic and excitonic states. Only the equilibrium positions $Q_k^0$ change for different states, namely $Q_k^{0,i} \neq Q_k^{0,j}$. We can define the dimensionless displacements $\Delta_k^{ij}$ as

$$\Delta_k^{ij} = \left(\frac{M_k \omega_k}{\hbar}\right)^{\frac{1}{2}} \left(Q_k^{0,i} - Q_k^{0,j}\right), \tag{29}$$

which measures the displacement along the kth normal mode when the electron changes from state i to state j.

Following Freed and Jortner,[55] the transition rate between two states through the multiple-phonon process is

$$\gamma_{i \to j}^{MP} = \sum_k \frac{\hbar^2}{M_k} \left|C_k^{ij}\right|^2 \hbar \omega_k \pi \frac{1}{\hbar D_{ij}^k \sqrt{2\pi}} \left( \begin{array}{c} (n_k + 1) \exp\left(-\frac{\left(\Delta E_{ij} - \hbar \omega_k - E_M^{ij}\right)^2}{2\hbar^2 \left(D_{ij}^k\right)^2}\right) \\ + n_k \exp\left(-\frac{\left(\Delta E_{ij} + \hbar \omega_k - E_M^{ij}\right)^2}{2\hbar^2 \left(D_{ij}^k\right)^2}\right) \end{array} \right) \tag{30}$$

with

$$\left(D_{ij}^k\right)^2 = \frac{1}{2} \sum_k \omega_k^2 \left(\Delta_k^{ij}\right)^2 (2n_k + 1)$$

$$n_k = \frac{1}{\exp(\hbar \omega_k / k_B T) - 1}$$

$$E_M^{ij} = \frac{1}{2} \sum_k \hbar \omega_k \left(\Delta_k^{ij}\right)^2$$

Herein the rearrangement energy $E_M^{ij}$ will be neglected without losing calculation accuracy, since $E_M^{ij}$ are usually very small. Unlike $\gamma_{i \to j}^{SP}$ with a Lorentzian lineshape, $\gamma_{i \to j}^{MP}$ exhibits a Gaussian lineshape, with its spectral width $D_{ij}$ temperature-dependent. Note that both $\gamma_{i \to j}^{SP}$ and $\gamma_{i \to j}^{MP}$ share the same form as the peak intensities go to infinity and the linewidths become $\delta$-functions, which is as follows

$$\gamma_{i \to j} = \sum_k \frac{\hbar^2}{M_k} \left|C_k^{ij}\right|^2 \hbar\omega_k \pi \left[ (n_k + 1) \delta(E_i - E_j - \omega_k) + n_k \delta(E_i - E_j + \omega_k) \right], \quad (31)$$

It is the decay mechanism that determines how the delta functions are broadened.

The total relaxation rates of an electronic state through electron-phonon interaction are expressed as the sum of all decay rates between this state and states with lower energies

$$\gamma_i^{E-P} = \sum_{j, |E^{0,j}| < |E^{0,i}|} \left( \gamma_{i \to j}^{SP} + \gamma_{i \to j}^{MP} \right) \quad (32)$$

## 3. NUMERICAL IMPLEMENTATIONS AND EXCITONIC NONRADIATIVE RELAXATION RATES

A smaller silicon cluster, $Si_{20}$, has been investigated and reported to demonstrate our methods developed for the electronic and excitonic inelastic scattering rates [39]. With the developed approach, we study the nonradiative relaxation rates of electrons and excitons in a larger cluster, $Si_{46}$. The NC (cluster) of $Si_{46}$ is selected for two reasons. First, we need a relatively larger nano-cluster to narrow the distribution of the data points, which will facilitate our analyses. Second, our previously reported cluster ($Si_{20}$) has degenerate states because of its $C_{3v}$ symmetry. $Si_{46}$ does not have any degenerate state because of its $C_{2v}$ symmetry. Thus we can focus on the nonradiative relaxation rates for $Si_{46}$.

The ground state LDA calculation is performed using the SIESTA code.[64] The core electrons $[1s^2 2s^2 2p^6]$ of Si are replaced by the nonlocal norm-conserving pseudopotential based on the Troullier-Martins scheme.[65] A quintuple-$\zeta$ double-polarization (5Z2P) basis set of numerical atomic orbitals is used for the four valence electrons of Si. The optimized structure of Si$_{46}$ is illustrated in Fig. 3, which has an $C_{2v}$ symmetry.

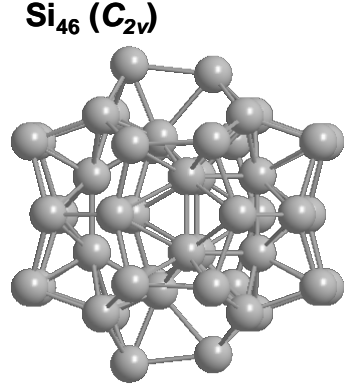

**Si$_{46}$ ($C_{2v}$)**

Fig. 3. Optimized structure of Si$_{46}$ with the $C_{2v}$ symmetry.

All integrals are evaluated on a uniform grid in real space with a grid spacing of 0.5 a.u.. The exchange integrals $\int d\mathbf{r} \int d\mathbf{r}' \varphi_i(\mathbf{r}) \varphi_j(\mathbf{r}) V(\mathbf{r},\mathbf{r}') \varphi_k(\mathbf{r}') \varphi_l(\mathbf{r}')$ are evaluated by first solving Poisson equations with the multigrid method.[66] The convergence of the QP calculation usually requires a large number of unoccupied states for the evaluation of the polarizability. Thus a Coulomb-hole screened-exchange (COHSEX) remainder scheme[48] has been applied to accelerate the convergence of the correlation part $\langle \varphi_i | \Sigma_c | \varphi_i \rangle$.

The properties of the one-particle states are obtained by solving the quasiparticle equation in section 2.1.1. After applying analytical continuation of $\Sigma_c(\mathbf{r},\mathbf{r}';E)$ in the complex energy plane, the energy $E_i$ and the inelastic scattering rates $\gamma_i$ of an electronic state are obtained by solving a set of complex equation set numerically (Eq. 8)[38]. The inelastic scattering rates of excitons are calculated with the method in sections 2.1.2. and 2.1.3. as developed in our previous article [39].

The normal coordinates $Q_k$ and frequencies $\omega_k$ are obtained by diagonalizing the mass-weighed second-order force matrix (Hessian matrix).[67] The second order derivatives $\partial^2 U/\partial X_i \partial X_j$ are calculated with a finite difference approach. Here $U$ is the total potential energy among all electrons and nuclei. $X_i$ are nuclear Cartesian coordinates. A unitary matrix without translational and rotational vectors is used to transform the Hessian matrix to a block diagonal matrix with each block corresponding to an irreducible representation[67]. Block off-diagonal elements are small and are thus eliminated to ensure that each $Q_k$ belongs to a specific irreducible representation exclusively.

The force on the $i^{th}$ atom due to the jth electronic state is calculated as $\mathbf{f}_i^j$ by our modified version of the SIESTA code. The calculation sums up all the energy derivatives associated with the $j^{th}$ electronic state, namely those from the kinetic energy, the non-local pseudopotential energy, Hartree energy, exchange-correlation energy and basis overlap.[64] Then the shift of the $i$th atom due to the $j$th electronic state is estimated as $\Delta X_{i,n}^j \approx \dfrac{f_{i,n}^j}{\partial^2 U / \left(\partial X_{i,n}^j\right)^2}$, $n = 1, 2, 3$

Therefore the shift along the $k^{th}$ normal coordinate due to the jth electronic state $\Delta Q_k^j$ can be obtained by the inner product between $\Delta \mathbf{X}^j$ and $\mathbf{Q}_k$, where $\mathbf{Q}_k$ is the vector representation of $Q_k$ in Cartesian coordinates. Here we do not take into account the Jahn-Teller effect, since the silicon cluster investigated here does not have degenerate electronic states. The pseudo-Jahn-Teller effect is not included either. Therefore we only need to calculate those $\Delta Q_k$ belonging to the irreducible representation with the total symmetry, namely $A_1$ of the $C_{2v}$ point group, since the square of each irreducible representation of the $C_{2v}$ point group is $A_1$.

The derivative $\left|\langle \Phi_j | \partial/\partial Q_k | \Phi_i \rangle\right|$ is evaluated by a finite difference method, which is more accurate than the frequently used perturbation method in literature. In the SIESTA code, the molecular orbitals are expressed as the linear combination of atomic orbital (LCAO),

$$\Phi_i(\mathbf{r}) = \sum_m c_{i,m} \phi_m(\mathbf{r}) \qquad (33)$$

where $\phi_m(\mathbf{r})$ is the $m^{\text{th}}$ atomic orbital. Then we have

$$\frac{\partial \Phi_i(\mathbf{r})}{\partial X} = \sum_m \frac{\partial c_{i,m}}{\partial X} \phi_m(\mathbf{r}) + \sum_m c_{i,m} \frac{\partial \phi_m(\mathbf{r})}{\partial X} \qquad (34)$$

and

$$\left\langle \Phi_j \left| \frac{\partial}{\partial X_k} \right| \Phi_i \right\rangle = \sum_{m,n} c_{j,n} \frac{\partial c_{i,m}}{\partial X_k} \int \phi_n(\mathbf{r}) \phi_m(\mathbf{r}) d\mathbf{r} + \sum_{m,n} c_{j,n} c_{i,m} \int \phi_n(\mathbf{r}) \frac{\partial \phi_m(\mathbf{r})}{\partial X_k} d\mathbf{r} \qquad (35)$$

Here we only take into account the internal conversion and neglect the intersystem crossing between singlet and triplet states arising from the spin-orbit coupling. The derivation and numerical treatment for excitonic states are similar to those of electronic states. First the force on the $i$th atom due to the $j$th excitonic state is calculated as

$$\mathbf{F}_i^j = \sum_{v,c} \left| R_{vc}^j \right|^2 \left( \mathbf{f}_i^c - \mathbf{f}_i^v \right) \qquad (36)$$

from which $\Delta X_{i,n}^j$, $\Delta Q_k^j$, $\Delta_k^{i,j}$ and $E_M^{i,j}$ can be obtained in exactly the same manner as in the case of electrons. The coupling term between the two excitonic states is approximated as

$$\left\langle \rho_j \left| \partial/\partial Q_k \right| \rho_i \right\rangle = \sum_{v,c} \sum_{v',c'} R_{vc}^i R_{v'c'}^j \left( \delta_{vv'} \left\langle \varphi_c \left| \partial/\partial Q_k \right| \varphi_{c'} \right\rangle + \delta_{cc'} \left\langle \varphi_v \left| \partial/\partial Q_k \right| \varphi_{v'} \right\rangle \right). \qquad (37)$$

Then Eqs. (27) and (30) can be extended to the calculation of excitonic nonradiative transition rates.

## 4. RESULTS AND DISCUSSIONS

### 4.1. Electronic Relaxation Dynamics in Si Clusters

The inelastic scattering rates $\gamma^{IS}$ of electrons and holes in the cluster $Si_{46}$ calculated by the *GW*Γ method are plotted versus the excitation energy $|E_i - E_F|$ in log-log style in Fig. 4. Note that all relaxation rates are presented in units of eV, which can be easily converted to fs$^{-1}$ by being divided by $\hbar = 0.658$ eV·fs. Our calculations show that the inelastic scattering rates of electrons and holes in Si$_{46}$ are similar to Si$_{20}$ which was presented in our previous article [39]. Specifically, the electrons and holes in Si$_{46}$ approach the quadratic law of Quinn and Ferell in the high-energy regime ($|E_i - E_F| > 6$ eV)

$$\tau_i^{IS} = 263 r_s^{-5/2} (E_i - E_F)^{-2} \text{ eV}^2 \text{ fs}. \tag{38}$$

where $\tau_i^{IS} = (2\gamma_i^{IS})^{-1}$ is the inelastic scattering lifetime.

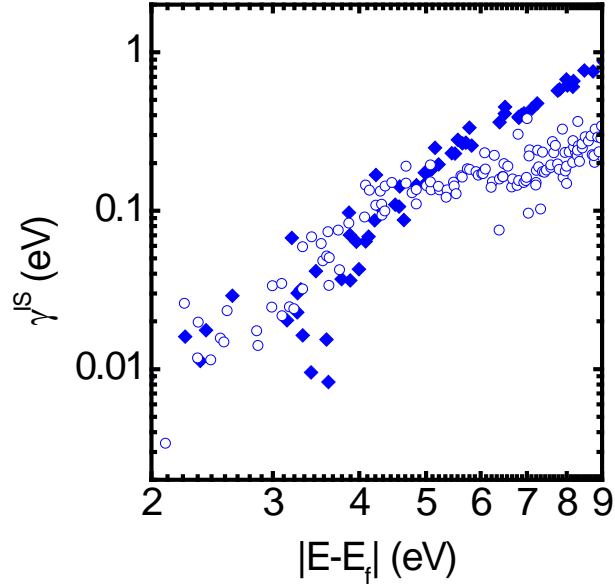

Fig. 4. Log-log plot of inelastic scattering rates $\gamma^{IS}$ of electrons (hollow circles) and holes (solid diamonds) in Si$_{46}$ vs energy from the Fermi level.

The single-phonon nonradiative relaxation rates $\gamma^{SP}$ of electrons and holes in Si$_{46}$ versus the excitation energy $|E_i - E_F|$ obtained at 0 K by Eq. (27) are plotted in Fig. 5a. Note that some data points in Fig. 4 do not have corresponding points in Fig. 5a, since $\gamma^{SP}$ for those states vanishes. This arises from the fact that the energy gaps between two neighboring states in a confined system may be larger than the maximum phonon frequency and thus the single-phonon relaxation mechanism between such two states is strictly prohibited. The multiple-phonon nonradiative relaxation rates $\gamma^{MP}$ at 0K in Si$_{46}$ are presented in Fig. 5b. The pattern of $\gamma^{MP}$ is quite dispersive. For some electron (or hole) states, the relaxation rates of the multiple-phonon process are even comparable to those of the single-phonon process. More importantly all electronic states (except the HOMO and LUMO) can decay through the multiple-phonon relaxation process, which is an alternative nonradiative decay pathway when the single-phonon process is absent. This shows that the multiple-phonon decay pathway is an important relaxation mechanism and should always be included for the study of nonradiative rates for finite systems.

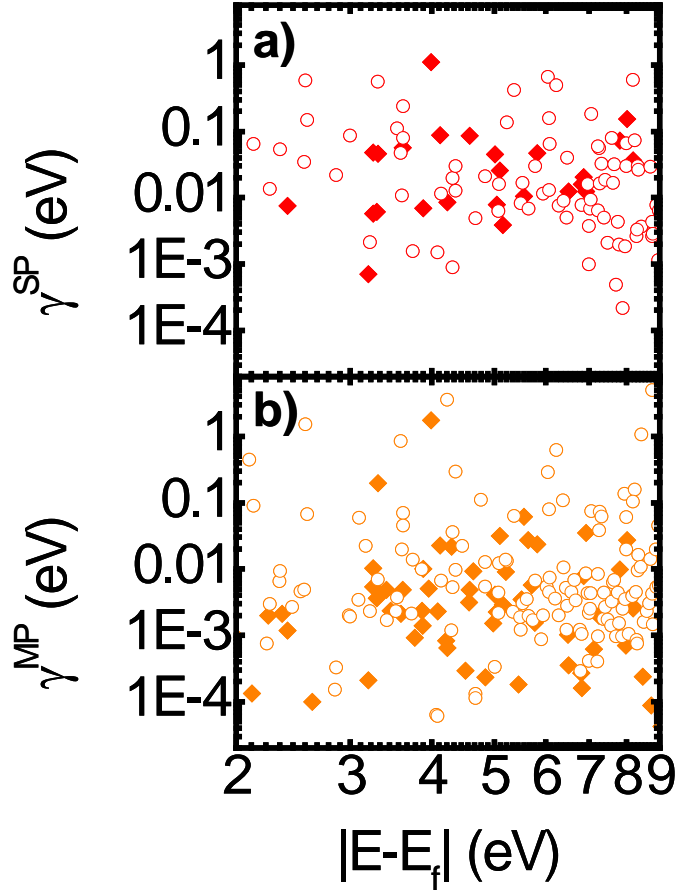

Fig. 5. Log-log plots of a) Single-phonon nonradiative relaxation rates $\gamma^{SP}$ and b) multiple-phonon nonradiative relaxation rates $\gamma^{MP}$ of electrons (hollow circles) and holes (solid diamonds) in $Si_{46}$ at 0 K.

The patterns of $\gamma^{SP}$ and $\gamma^{MP}$ in Figs. 5a and 5b are more dispersive than that of $\gamma^{IS}$ in Fig. 4. The reason is that the single-phonon relaxation can only occur between a state and those states with energies below it yet not too far away (within the phonon energy $\hbar\omega_k$). Therefore an electron in a given electronic state may have many strongly coupled states available for the single-phonon relaxation and thus exhibit a large $\gamma^{SP}$. On the hand, it may only have one or two weakly coupled states and thus present a small $\gamma^{SP}$. It means that $\gamma^{SP}$ is essentially a local quantity in terms of energy and is dependent upon the case being studied. This explains why the pattern of $\gamma^{SP}$ is so dispersive and even two states with close excitation energies $|E_i - E_F|$ may have quite different $\gamma^{SP}$. Note that although the multiple-phonon relaxation can occur in principle between a state and any states with energies far below it, in practice there is still an upper limit for the energy gaps, as $\gamma^{MP}$ decreases exponentially with increasing energy gap. It means that $\gamma^{SP}$ is a semi-local quantity in terms of energy and is also dependent upon the case being studied. Therefore the same interpretation applies to $\gamma^{MP}$. The local and semi-local properties of single-phonon and multiple-phonon relaxation processes distinguish notably from the inelastic scattering relaxation, where an electron in an electronic state can transit to those states with energies far below the initial state, namely no upper limit for energy gaps. Therefore, the higher the energy of the initial state, the larger the inelastic scattering rate is. In this case, the absolute excitation energy $|E_i - E_F|$ does matter.

The ratios $\gamma^{IS}/\gamma^{SP+MP}$ ($\gamma^{SP+MP} = \gamma^{SP} + \gamma^{MP}$) are plotted in Fig. 6 for the comparison of the inelastic scattering rates and the nonradiative relaxation rates of electronic states in $Si_{46}$. The patterns for $\gamma^{IS}/\gamma^{SP+MP}$ are even more dispersive than that of $\gamma^{MP}$, with some data points above unity and the others below. This implies that inelastic scattering is highly possible to occur for some electronic states, while nonradiative relaxation will dominate the other electronic states. However, the data in Fig. 6 suggest that the inelastic scattering relaxation is a more significant effect, since the logarithmic mean of data in Fig. 6 are larger than unity. In addition, nonradiative relaxation progresses in a cascade manner. Consequently, it is highly possible that an electronic state with high excitation energy passes through several intermediate electronic states during its nonradiative relaxation. The inelastic scattering decay will occur sooner or later, as long as one of these intermediate states favors inelastic scattering more. Therefore one can conclude that inelastic scattering (or reverse Auger / impact ionization) can always occur for electronic states with high excitation energy.

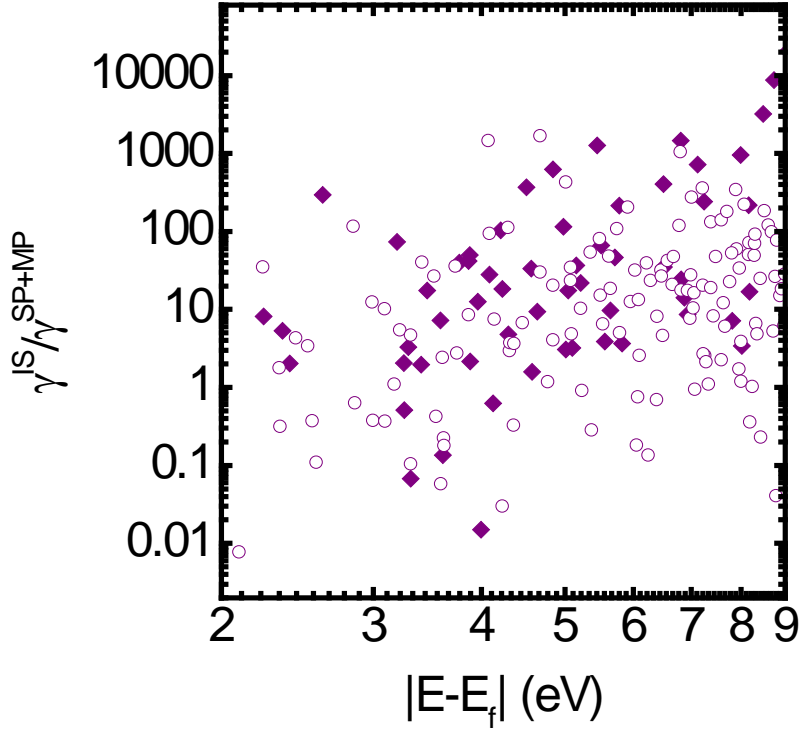

Fig. 6. Log-log plot of the ratios $\gamma^{IS}/\gamma^{SP+MP}$ for electrons (hollow circles) and holes (solid diamonds) in $Si_{46}$ at 0 K, where $\gamma^{SP+MP} = \gamma^{SP} + \gamma^{MP}$.

The temperature effect is studied by re-calculating $\gamma^{SP}$, $\gamma^{MP}$ and $\gamma^{IS}/\gamma^{SP+MP}$ at 300 K. The results are illustrated in Figs. 7 and 8. Here we assume that the electronic scattering rates are temperature-independent. As the temperature changes from 0 to 300 K, all $\gamma^{SP}$ and $\gamma^{MP}$ are enhanced with a factor ranging from 1 to 3. Usually the temperature enhancement of $\gamma^{MP}$ is larger than the corresponding enhancement of $\gamma^{SP}$, since the increased temperature not only elevates the average quantum number $n_k$ of each normal mode for both $\gamma^{SP}$ and $\gamma^{MP}$, but also increases the thermal broadening factor $D_{ij}^k$ in Eq. (30) solely for $\gamma^{MP}$. Results in Fig. 8 show that the ratios of $\gamma^{IS}/\gamma^{SP+MP}$ are reduced at 300K compared to the ratios of $\gamma^{IS}/\gamma^{SP+MP}$ at 0K as shown in Fig. 6. Yet it is still essential to include the multiple-phonon mechanism for calculations of $\gamma^{IS}/\gamma^{SP+MP}$, $\gamma^{IS}$ and $\gamma^{SP+MP}$.

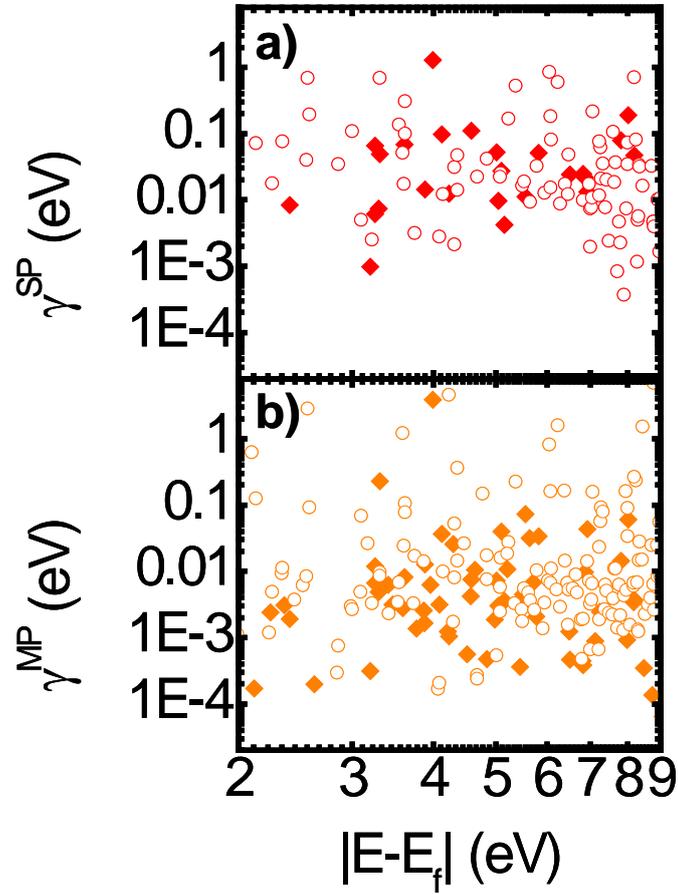

Fig. 7. Log-log plots of a) Single-phonon nonradiative relaxation rates $\gamma^{SP}$ and b) multiple-phonon nonradiative relaxation rates $\gamma^{MP}$ of electrons (hollow circles) and holes (solid diamonds) in $Si_{46}$ at 300 K.

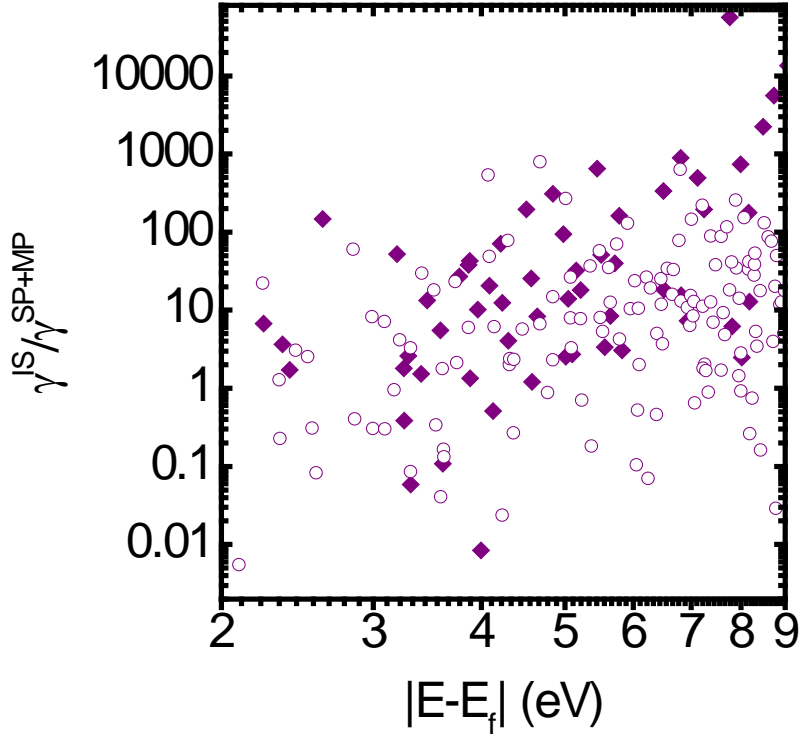

Fig. 8. Log-log plot of the ratios $\gamma^{IS}/\gamma^{SP+MP}$ for electrons (hollow circles) and holes (solid diamonds) in $Si_{46}$ at 300 K, where $\gamma^{SP+MP} = \gamma^{SP} + \gamma^{MP}$.

It is known that there are different possible ways to excite a multi-electron system. One possible way is to add an electron to the system, or remove an electron from it to excite a hole. These excitations exactly correspond to the electronic states discussed here. It should be pointed out that these electronic states are essentially charge-non-conserved one-particle excitations, which are not directly involved in most optical applications of semiconductor nanoclusters. Actually in photovoltaic systems based on semiconductor nanoclusters, most incident photons just induce charge-conserved excitonic (electron-hole) excitations. Therefore the relaxation dynamics of excitons is of greater importance, which will be addressed in the next section.

**4.2. Excitonic Relaxation Dynamics in Si Clusters**

The inelastic scattering rates $\Gamma^{IS}$ (capital letter stands for excitons) of excitons in $Si_{46}$ are plotted versus the excitation energy $\Omega$ in log-log style in Fig. 9, where the solid line is the curve fitting of $\Gamma^{IS}$ by a simple rational function (Padé function $P_1^2$)

$$y^{IS} = 2x + a + \frac{b}{x+c} \qquad (39)$$

where $x$ and $y$ represent $\ln(\Omega/\text{eV})$ and $\ln(\Gamma^{IS}/\text{eV})$, respectively. The fitting coefficients a, b and c are -5.00, -0.22 and -0.40, respectively. The factor of the linear term in $x$ is fixed to be 2, since it is easy to prove that the quadratic relation between the excitonic decay rate and the excitonic energy will be approached at the high-energy limit (large x), provided that the quadratic relation between the QP decay rate and the QP energy is approached at the high energy regime.

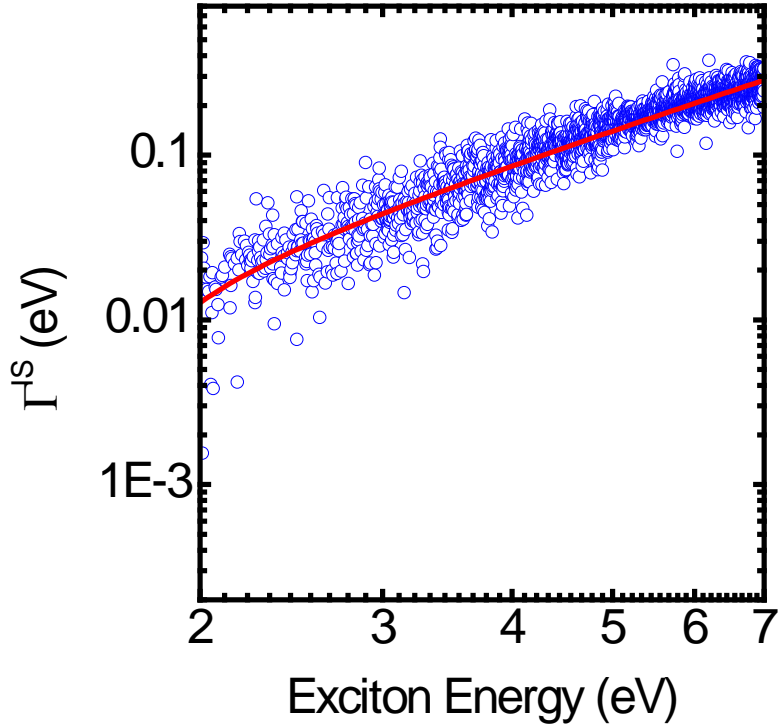

Fig.9. Log-log plot of the inelastic scattering rates $\Gamma^{IS}$ of excitons in $Si_{46}$ vs exciton energy. The solid line is the curve-fitting.

The single-phonon nonradiative relaxation rates $\Gamma^{SP}$ of excitons in $Si_{46}$ at 0 K obtained by Eq. (27) are plotted versus the excitation energy $\Omega_s$ in Fig. 10a, where the $\Gamma^{SP}$ points are found to be energy-dependent in the low-energy regime. Unlike $\Gamma^{IS}$ shown in Fig. 9, such an energy-dependence diminishes in the high-energy regime and the pattern of $\Gamma^{SP}$ becomes flat, although the data distribution is still wide. This is attributed to the large excitonic density of state (DOS) that quickly saturates the exciton-phonon interaction. According to the pattern shown in Fig. 10a, we fit the data with an exponential function

$$y^{SP} = y_0^{SP} + A^{SP} e^{-(\Omega - \Omega_0^{SP})/t^{SP}} \tag{40a}$$

where $y^{SP}$ represents $\log(\Gamma^{SP}/\text{eV})$, $\Omega$ is the excitonic energy, and $y_0^{SP}$, $A^{SP}$, $\Omega_0^{SP}$ and $t^{SP}$ are fitting parameters. The fitting curve is plotted in Fig. 9a as a solid line. Here $y_0^{SP} = -0.826$, which leads to the converged $\Gamma^{SP}$ that is estimated to be 0.149 eV.

The multiple-phonon nonradiative relaxation rates $\Gamma^{MP}$ of excitons in Si$_{46}$ at 0 K obtained by Eq. (30) are presented in Fig. 10b. It can be seen that the pattern of $\Gamma^{MP}$ is similar to that of $\Gamma^{SP}$ as shown in Fig. 10a. Thus the data are fitted with the same equation

$$y^{MP} = y_0^{MP} + A^{MP} e^{-(\Omega - \Omega_0^{MP})/t^{MP}} \tag{40b}$$

Here $y_0^{MP} = -0.767$, which yields the converged $\Gamma^{MP}$ to be 0.170 eV. Note that $\Gamma^{MP}$ are always comparable to $\Gamma^{SP}$ in the full energy range studied. This again demonstrates the necessity to include the multiple-phonon decay mechanism for the simulation of nonradiative relaxation rates. Furthermore, both $\Gamma^{SP}$ and $\Gamma^{MP}$ in the cluster range from 0.1 to 1000 meV, which correspond to nonradiative relaxation lifetimes ranging from several picoseconds to about a femtosecond. Such a fast nonradiative relaxation process implies that the phonon bottleneck does not apply in the present structure Si$_{46}$. Some previous researchers have predicted multiphonon emissions in CdSe and PbSe NCs[68]. Ref. 7 reported a nonradiative relaxation time between 10fs and 500fs for 3.9nm PbSe NCs. According to ref. 69, the nonradiative relaxation in PbSe NCs should be by multiphonon emission. Though the NCs in the studies are different from our NC studied (Si46), it is still useful to qualitatively compare these results with our modeling: first, our computation proves that multiphonon emission is indeed important in NCs; second, our computed $\Gamma^{SP}$ and $\Gamma^{MP}$ (1fs ~ 1000fs) for Si NC is the range of these results.

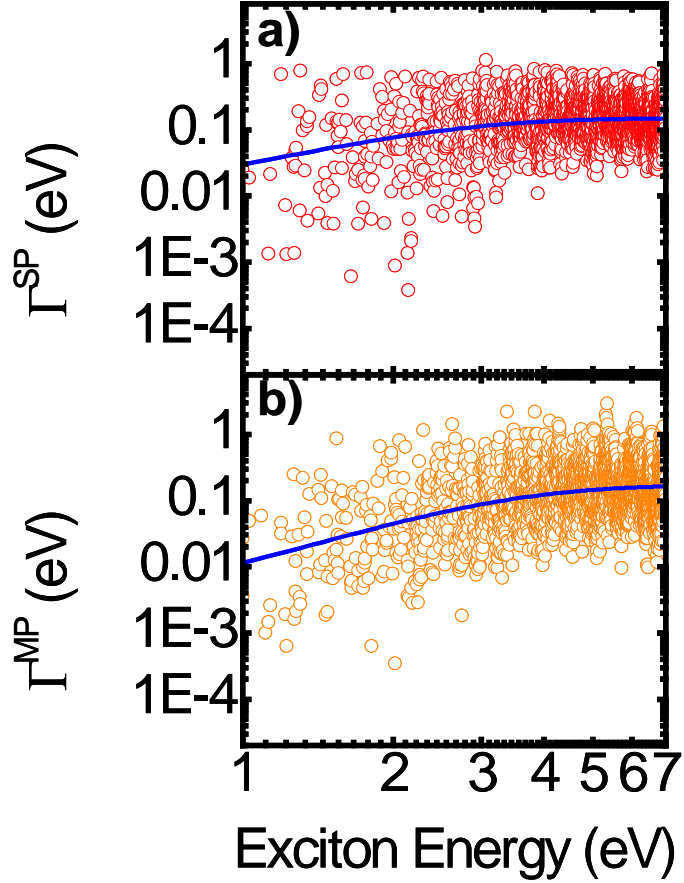

Fig.10.  a) Single-phonon nonradiative relaxation rates $\Gamma^{SP}$ and b) multiple-phonon nonradiative relaxation rates $\Gamma^{MP}$ of excitons in $Si_{46}$ at 0 K. Solid lines are curve fitting based on Eq. (40).

The ratios $\Gamma^{IS}/\Gamma^{SP+MP}$ ( $\Gamma^{SP+MP} = \Gamma^{SP} + \Gamma^{MP}$ ) are plotted in Fig. 11 for the comparison of the inelastic scattering rates and the nonradiative relaxation rates of excitonic states in $Si_{46}$. The ratio $\Gamma^{IS}/\Gamma^{SP+MP}$ in Fig. 11 increases steadily with increasing excitonic energy. It is consistent with the fact that $\Gamma^{IS}$ increases almost quadratically with increasing excitonic energy (Fig. 9), while $\Gamma^{SP}$ and $\Gamma^{MP}$ approach a constant in the high-energy regime (Fig. 10). It should be emphasized here that most ratios $\Gamma^{IS}/\Gamma^{SP+MP}$ for excitons are smaller than unity, which notably differs from $\gamma^{IS}/\gamma^{SP+MP}$ for electronic states (Fig.6). This again can be attributed to large excitonic DOS and will be addressed in the next section.

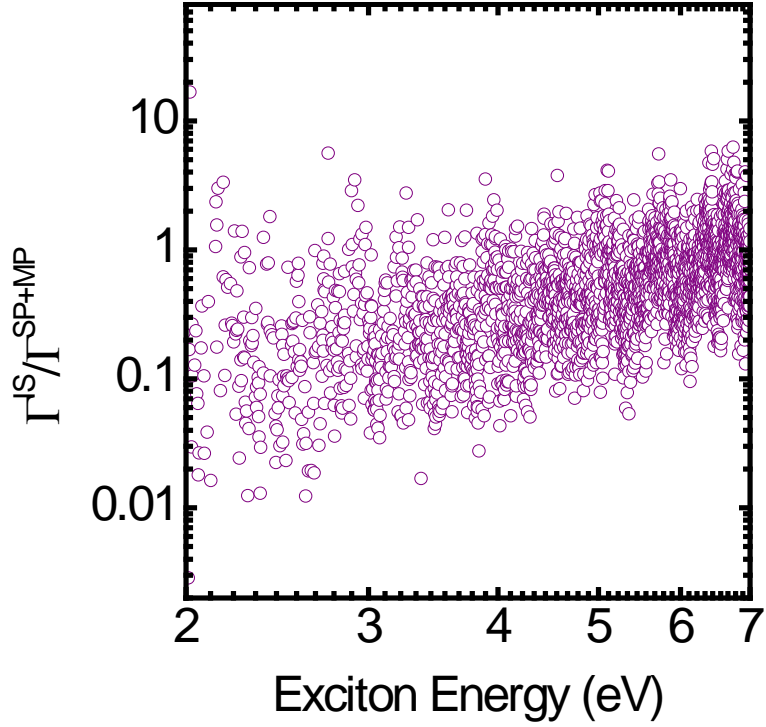

Fig.11. The ratios $\Gamma^{IS}/\Gamma^{SP+MP}$ for excitons in Si$_{46}$ at 0 K, where $\Gamma^{SP+MP} = \Gamma^{SP} + \Gamma^{MP}$.

The temperature effect is investigated by re-calculating $\Gamma^{SP}$, $\Gamma^{MP}$ and $\Gamma^{IS}/\Gamma^{SP+MP}$ at various temperatures. In this way, the influence of the temperature on the nonradiative relaxation of excitons can be tracked in a way more quantitative, since now we can characterize $\Gamma^{SP}$ and $\Gamma^{MP}$ with fitting coefficients based on Eq. (40). In most cases we are interested in $\exp(y_0^{SP})$ and $\exp(y_0^{MP})$, namely the converged relaxation rates of single-phonon and multiple-phonon processes. The ratios of $\exp(y_0^{SP})_T / \exp(y_0^{SP})_{T=0}$ and $\exp(y_0^{MP})_T / \exp(y_0^{MP})_{T=0}$ over the full temperature range and the low temperature range are plotted versus temperature in Figs. 12a and 12b, respectively. Here $\exp(y_0^{SP})_{T=0}$ and $\exp(y_0^{MP})_{T=0}$ have been used as the units for the two data sets, respectively.

Assume that all phonon frequencies can be represented by a characteristic phonon frequency $\bar{\omega}^{SP}$ for the single-phonon relaxation process. According to Eq. (27), we can obtain

$$\exp\left(y_0^{SP}\right)_T \Big/ \exp\left(y_0^{SP}\right)_{T=0} \approx 2\bar{n}^{SP} + 1 = \frac{2}{\exp\left(\hbar\bar{\omega}^{SP}/k_BT\right)-1} + 1 \approx \frac{2k_BT}{\hbar\bar{\omega}^{SP}} + C^{SP} \quad (41a)$$

where $\bar{n}^{SP}$ is the average quantum number of the characteristic phonon frequency $\bar{\omega}^{SP}$. The last approximate equality in Eq. (42a) is valid only in the high temperature limit. A similar equation can be written to define the characteristic phonon frequency $\bar{\omega}^{MP}$ for the multiple-phonon relaxation process.

$$\exp\left(y_0^{MP}\right)_T \Big/ \exp\left(y_0^{MP}\right)_{T=0} \approx 2\bar{n}^{MP} + 1 = \frac{2}{\exp\left(\hbar\bar{\omega}^{MP}/k_BT\right)-1} + 1 \approx \frac{2k_BT}{\hbar\bar{\omega}^{MP}} + C^{MP} \quad (41b)$$

Note that Eq. (41b) is an approximation rougher than Eq. (41a), since the temperature effect on $D_{ij}^k$ in Eq. (30) has been neglected.

The two data sets as shown in Fig. 12a differ from the shape of the function $f = \frac{2}{\exp\left(\hbar\omega_k/k_BT\right)-1} + 1$ for a single phonon with the frequency $\omega_k$. Nevertheless, we can still perform the linear fitting in the temperature range of interest, as shown in Fig. 12b. According to the slopes of the two fitting lines in Fig. 12b, two characteristic phonon frequencies $\bar{\omega}^{SP}$ and $\bar{\omega}^{MP}$ for single-phonon and multiple-phonon relaxation processes can be solved based on Eqs. (41a) and (41b). Numerically, $\bar{\omega}^{SP}$ and $\bar{\omega}^{MP}$ are just the inverse of the weighted average of the inverse phonon frequencies, and thus have some important physical information over the selected temperature range for the cluster investigated.

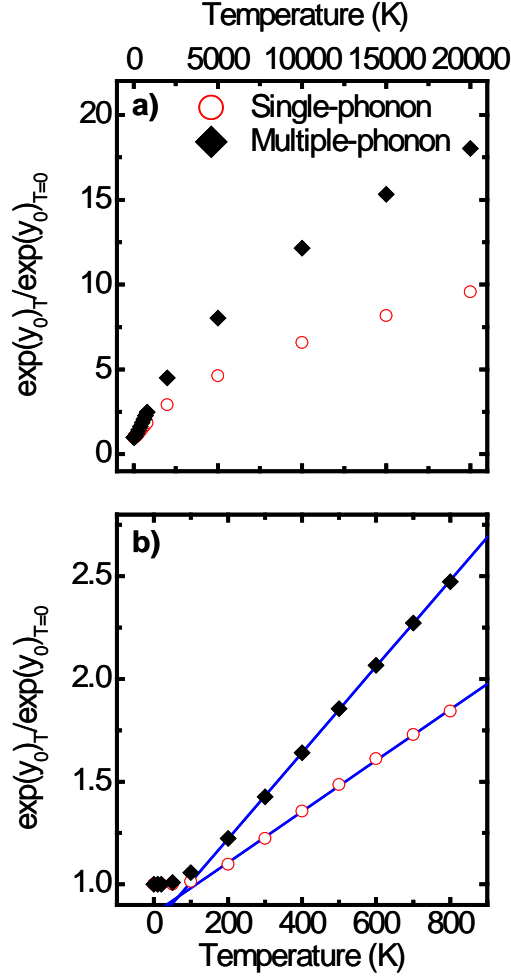

Fig.12. The ratios $\exp(y_0^{SP})_T / \exp(y_0^{SP})_{T=0}$ and $\exp(y_0^{MP})_T / \exp(y_0^{MP})_{T=0}$ versus temperature for $Si_{46}$ a) over the full temperature range and b) over the low temperature range.

From the slopes of the two linear lines in Fig. 12b, $\bar{\omega}^{SP}$ and $\bar{\omega}^{MP}$ are estimated to be 280 and 165 cm$^{-1}$. The phonon density of state (DOS) of $Si_{46}$ is plotted vs wavenumber in Fig. 13, where the locations of $\bar{\omega}^{SP}$ and $\bar{\omega}^{MP}$ are also given as dashed lines. It is found that $\bar{\omega}^{SP}$ is located at the middle of the phonon DOS spectrum. This implies that both the low- and high-frequency phonons contribute almost equally to the single-phonon relaxation process. On the other hand, $\bar{\omega}^{MP}$ emerges at the low-frequency regime. We thus speculate that the low-frequency phonons make more contribution to the multiple-phonon relaxation than the high-frequency phonons.

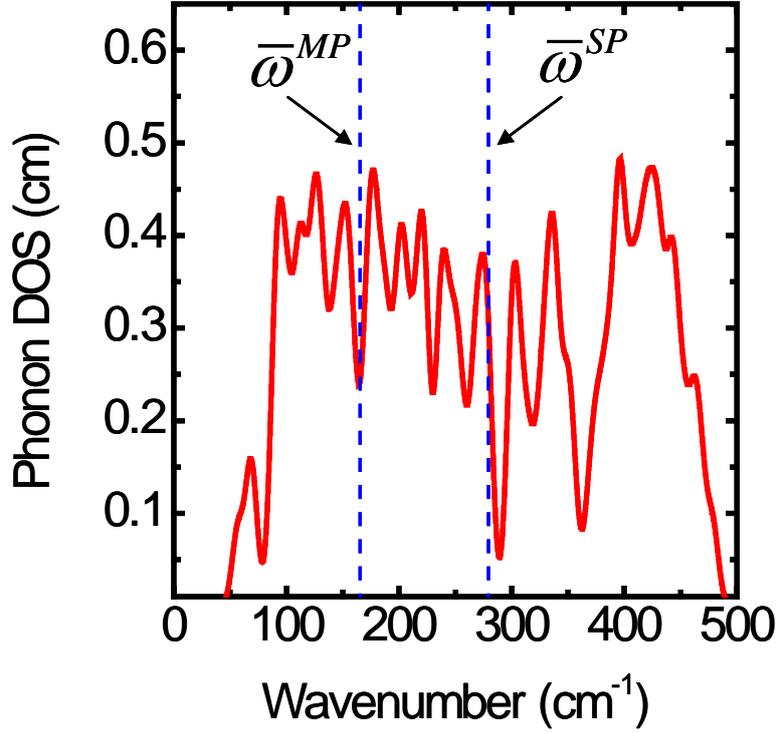

Fig.13. The phonon density of state (DOS) of $Si_{46}$, the locations of $\bar{\omega}^{SP}$ and $\bar{\omega}^{MP}$ are indicated as dashed lines.

**4.3. Comparison between the Electronic and Excitonic Nonradiative Relaxation Dynamics**

The nonradiative relaxation dynamics of electronic states and excitonic states discussed in the two previous sections have demonstrated notable differences as can be seen by comparing Figs. 5 and 6 with Figs. 10 and 11. The observations are attributed to the difference between the electronic DOS and the excitonic DOS illustrated in Figs. 14a and 14b. It can be seen from Fig. 14 that the electronic DOS is in the order of tens per eV, while the excitonic DOS is around several hundred per eV, namely one order of magnitude larger than the electronic DOS, and the excitonic DOS increases almost linearly with increasing exciton energy. The difference between the two types of DOS can be interpreted with a simple model, where the DOS of electrons is $g_c(E)$, and DOS of holes is $g_v(E)$. Then the DOS of excitons $g_{exc}(E)$ can be expressed approximately as a convolution of $g_c(E)$ and $g_v(E)$

$$g_{exc}(E) \approx \int dE' g_c(E'+E) g_v(E') \tag{42}$$

Eq. (42) shows $g_{exc}(E)$ is a function that is one order or magnitude higher than $g_c(E)$ and $g_v(E)$, since the former is generated through the product of the latter two. As a convolution, $g_{exc}(E)$ is also much smoother than $g_c(E)$ and $g_v(E)$. As long as $g_c(E)$ and $g_v(E)$ do not vary too dramatically in the energy range studied, we may further simply them as

$$g_c(E) = \begin{cases} C_1 & (E > 0) \\ 0 & (E < 0) \end{cases} \tag{43a}$$

$$g_v(E) = \begin{cases} 0 & (E > 0) \\ C_2 & (E < 0) \end{cases} \tag{43b}$$

Thus the convolution of $g_c(E)$ and $g_v(E)$ becomes a linear function $C_1 C_2 E$, which also explains the quasi-linear relation between $g_{exc}(E)$ and $E$ shown in Fig. 14b.

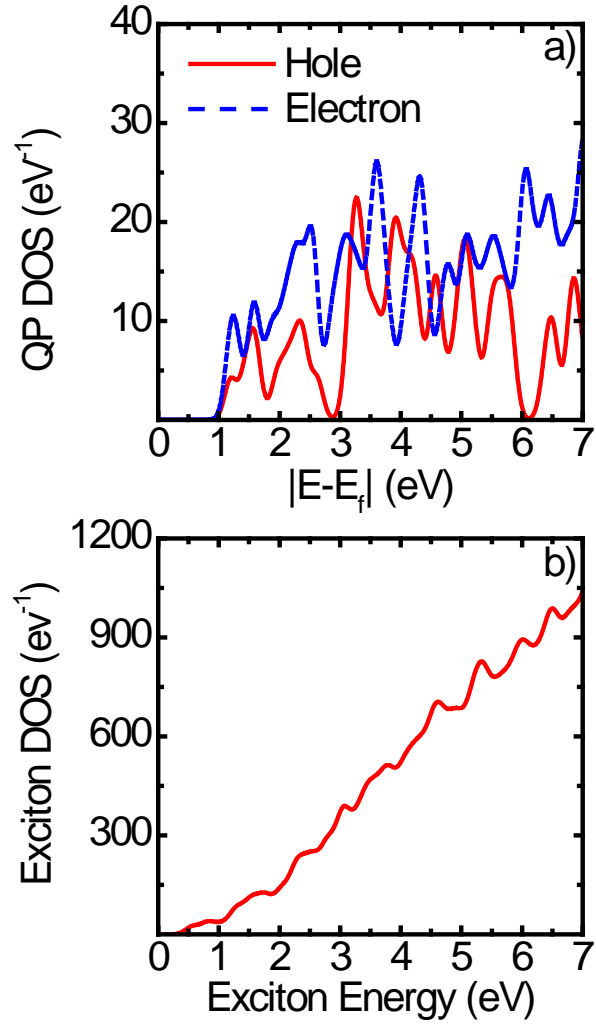

Fig.14. a) Electronic density of state (DOS) vs energy deference from the Fermi level and b) Excitonic DOS of $Si_{46}$ vs exciton energy.

With the understanding of the difference between the electronic DOS and excitonic DOS, we revisit Fig. 5 and Fig. 10. In the case of nonradiative relaxation of electrons and holes (Fig. 5), the electronic DOS is not large enough to saturate every electronic state. (Here "saturate" means to provide all possible final states for the nonradiative decay of a given initial state.) Therefore, some states may present relatively low single-phonon or multiple-phonon relaxation rates due to lack of decay pathways, which may occur even for those states with large excitation energies. In the case of nonradiative relaxation of excitons, however, the excitonic DOS increases quickly with increasing excitation energy and saturates those high-energy excitons very effectively. In summary, the calculation and analysis in the study uncover an important discovery: The nonradiative relaxation of excitons should always be investigated within the two-particle framework, otherwise the underlying physics will be missed and suspicious or plausible computational results will be presented.

## 5. CONCLUSION

. We have established the computational methodology for electron-phonon and exciton-phonon interaction in nanoclusters. Using the method, we have investigated the dynamics of electrons and excitons in a silicon cluster of $Si_{46}$. The nonradiative relaxation rates of electrons and excitons are calculated. It is found that the single-phonon and multiple-phonon relaxation mechanisms should be studied separately, as the two mechanisms correspond to two types of physical processes and have totally different spectral lineshapes. The single-phonon relaxation mechanism shall correspond to a Lorentzian function, which can be accounted for by the imaginary part of an electron-phonon self-energy term. The multiple-phonon relaxation mechanism, on the other hand, shall be related to a Gaussian function, which corresponds to the thermal process and involves the anharmonic effect (displacement of the potential energy surface in finite systems). It is also demonstrated that the formula derived for the two relaxation mechanisms share a general form at the delta-function limit.

An energy criterion distinguishing the single-phonon relaxation and multiple-phonon relaxation has been proposed for practical implementation of the computation. Our numerical results show that the multiple-phonon relaxation always exist and its rates are comparable to the corresponding single-phonon relaxation rates, for both electrons and excitons in the system studied ($Si_{46}$). It is thus essential to include the multiple-phonon relaxation mechanism when studying the nonradiative relaxation in small systems such as semiconductor nanoclusters.

Another important argument is that the nonradiative relaxation of electronic states and excitonic states should always be distinguished. The large difference between the DOS of exciton and electrons, contributes to a significant difference between the relaxations dynamics of electrons and excitons. Electronic states, even those with high excitation energy, may present relatively slow nonradiative relaxation rates due to the lack of final states available for the decay transitions. For excitonic states, however, the nonradiative relaxation rate increases and converges quickly with increasing exciton energy, due to the large excitonic DOS.

The temperature effect of the nonradiative relaxation of excitons in $Si_{46}$ has been investigated quantitatively. According to the average phonon frequencies derived from the data in the high-temperature regime, we speculate that both high- and low-frequency phonons contribute almost equally to the single-phonon relaxation pathway of excitons in $Si_{46}$, while low-frequency phonons are the major sources for the multiple-phonon relaxation mechanism.

The inelastic scattering rates of electrons and excitons are calculated based on many-body Green's function theory. These results are also compared with the corresponding nonradiative relaxation rates. For the electronic states in $Si_{46}$, the inelastic scattering decay is predicted to be a major decay mechanism, and nonradiative relaxation rates are larger than inelastic scattering rates for most excitonic states in $Si_{46}$.

■AUTHOR INFORMATION

**Corresponding Author**

*E-mail: taofang@alum.mit.edu

**ORCID**

**Taofang Zeng : 0000-0001-6589-4621**

**Author Contributions**

The manuscript was written through contributions of all authors. All authors have given approval to the final version of the manuscript.

**Funding**

National Science Foundation (Grant No.CBET-0830098)

**Notes**

The authors declare no competing financial interest.

ACKNOWLEDGMENT

This project is sponsored by the National Science Foundation (Grant No.CBET-0830098) and Hunan provincial grant.